\documentclass[a4paper]{ISOStyle}
\usepackage{epsfig}
\usepackage{graphics}

\begin{document}

\title{IRS: the Infrared Spectrograph on SIRTF}

\author{J. Houck\inst{1} 
	\and J. van Cleve\inst{1} 
	\and B. Brandl\inst{1} 
	\and V. Charmandaris\inst{1} 
	\and D. Devost\inst{1} 
	\and K. Uchida\inst{1}} 
\institute{IRS Science Center, Center for Radiophysics \& Space Research, Cornell University, Ithaca NY 14853, USA}

\maketitle 

\begin{abstract}

The Infrared Spectrograph (IRS)\index{IRS} is one of the three
instruments on board the Space Infrared Telescope Facility
(SIRTF)\index{SIRTF} to be launched in December 2001. The IRS will
provide high resolution spectra (R$\sim$600) from 10--37\,$\mu$m
and low resolution spectra (R$\ge$60) from 5.3--40\,$\mu$m.  Its high
sensitivity and ``spectral mapping-mode'' make it a powerful
instrument for observing both faint point-like and extended sources.

\end{abstract}

\section{The Infrared Spectrograph}

\begin{figure}[!ht]
\centering
\includegraphics[width=0.48\textwidth]{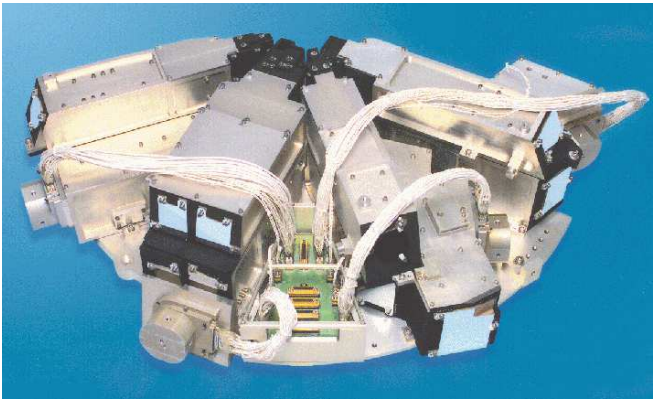}
\caption[]{The 4 IRS modules on their common base plate.}
\label{irs}
\end{figure}

The Infrared Spectrograph (IRS) (\cite{houck}) will provide the Space
Infrared Telescope Facility (SIRTF) (\cite{fanson}) with low and
moderate-spectral resolution spectroscopic capabilities from 5.3 to 40
microns.  The IRS (see Fig.~\ref{irs}) is composed of four separate
modules, with two of the modules providing R$>$60 spectral resolution
over 5.3 to 40 microns and two modules providing R$\sim$600 spectral
resolution over 10 to 37 microns.

The IRS instrument has no moving parts (``bolt-and-go'' philosophy).
Each module has its own entrance slit in the focal plane.  The
low-resolution modules employ long slit designs that allow both
spectral and one-dimensional spatial information to be acquired
simultaneously on the same detector array. Two small imaging
sub-arrays (``peak-up cameras'') in the short-low module (SL) will
also allow infrared objects with poorly known positions to be
accurately placed into any of the IRS modules' entrance slits. The
high-resolution modules use a cross-dispersed echelle design that
gives both spectral and spatial measurements on the same detector
array.

\begin{figure}[!ht]
\centering
\includegraphics[width=0.48\textwidth]{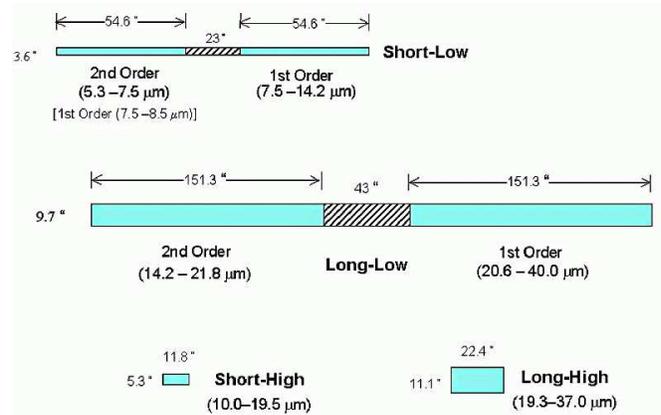}
\caption[]{Details on the IRS slits}
\label{slits}
\end{figure}

\begin{figure}[!ht]
\centering
\includegraphics[width=0.48\textwidth]{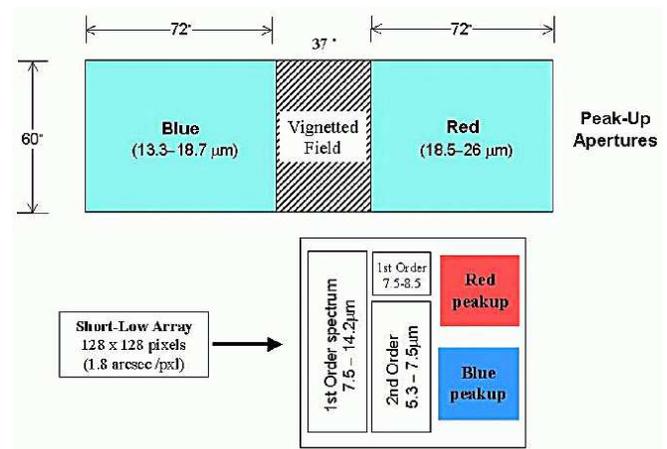}
\caption[]{The IRS peakup camera}
\label{peakup}
\end{figure}

\section{The IRS Sensitivity}

The expected sensitivity of IRS is much higher than the one of the
Infrared Space Observatory. The theoretical sensitivity plots for the
four modules are presented in following figures.

\begin{figure}[!ht]
\centering
\includegraphics[width=0.48\textwidth]{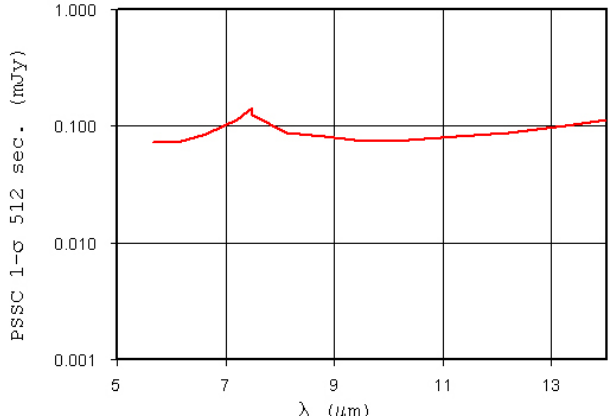}
\caption[]{The IRS Short-Low Point Source Staring Continuum sensitivity}
\end{figure}

\begin{figure}[!h]
\centering
\includegraphics[width=0.48\textwidth]{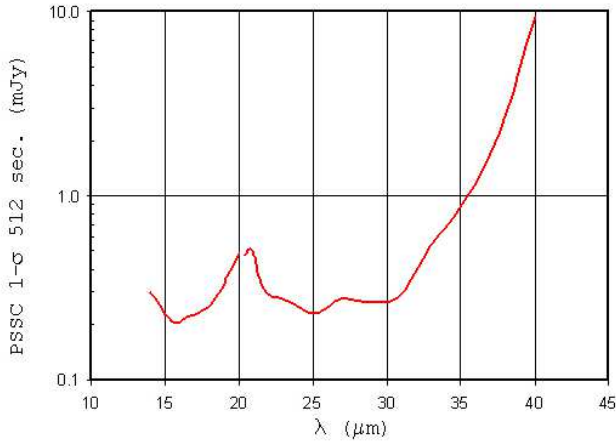}
\caption[]{The IRS Long-Low Point Source Staring Continuum sensitivity}
\end{figure}

\begin{figure}[!h]
\centering
\includegraphics[width=0.48\textwidth]{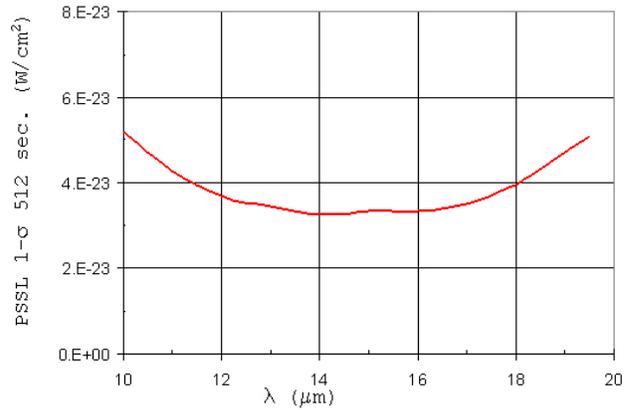}
\caption[]{{\it The IRS Short-High Point Source Staring Line sensitivity}}
\end{figure}

\begin{figure}[!h]
\centering
\includegraphics[width=0.48\textwidth]{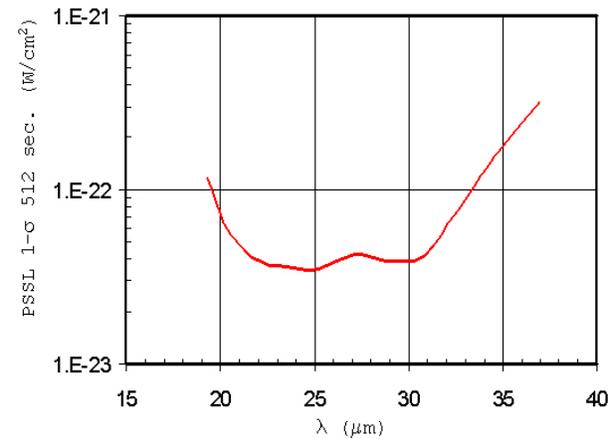}
\caption[]{{\it The IRS Long-High Point Source Staring Line sensitivity}}
\end{figure}

The continuum point source (7$\sigma$, 500 seconds) sensitivity of the
low resolution module is 1~mJy at 12\,$\mu $m.  The line sensitivity
of the high resolution module is 4$\times$10$^{-18}$ Wm$^{-2}$ at
15\,$\mu $m. The two peak-up cameras can center on sources as faint as
0.7~mJy.

The saturation limits in 8 seconds for point (extended) sources are 5~Jy
(0.4~Jy\,arcsec$^{-2}$) at 10\,$\mu $m for the low resolution module and
50~Jy (2.1~Jy arcsec$^{-2}$) at 15\,$\mu $m for the high resolution
module. The 4~s saturation limits for the peak-up cameras are 0.5~Jy for 
point sources and 40~mJy\,arcsec$^{-2}$ for extended sources.

\section{Discussion}

\begin{table*}[!ht]
\centering
\caption{Basic IRS characteristics}
\renewcommand{\arraystretch}{1.4}
\setlength\tabcolsep{5pt}
\begin{tabular}{cccccc}
\hline\noalign{\smallskip}
Module & Detector & Pixel Size & Slit Size & $\lambda$ & Resolving \\
& (128$\times$128) & (arcsec) & (arcsec) & ($\mu$m) &  Power (R)\\
\noalign{\smallskip}
\hline
\noalign{\smallskip}
Short Low  & Si:As & 1.8 & 3.6$\times$54.5 & 5.3 -- 7.5 & 62-124\\
 ''        &   ''    &  ''   &    ''             & 7.5 -- 14  &     ''  \\
Long Low   & Si:Sb & 4.8 & 9.7$\times$151.3 & 14 -- 21  & 62-124\\
''         &   ''    &   ''  &      ''            & 21 -- 40  &     ''  \\
Short High & Si:As & 2.4 & 5.3$\times$11.8  & 10 -- 19.5 & 600\\
Long High  & Si:Sb & 4.8 & 11.1$\times$22.4  & 19 -- 37  & 600\\
\hline
\end{tabular}
\label{Table1}
\end{table*}

The diversity of the mid-infrared spectral features, became evident in
this meeting. The intensity of the Unidentified Infrared Bands was
shown to vary substantially as a function of the intensity of the
radiation field. Their nature (molecules in aromatic bond structures)
and physical properties are still under debate, while new features, such
as the the 16.4 $\mu$m (\cite{verstraete}) are being discovered. It
appears though, that the carriers of these bands are present in the
photo-dissociation regions and their intensity reveals the presence of
star formation activity, often obscured in the optical wavelengths.

In conjunction with the other two instruments of SIRTF, the
mid-infrared camera (IRAC) and the mid/far-infrared photometer (MIPS),
IRS provides unique opportunities for a wide range of research
projects. It's worth mentioning that most of the sources detected by
IRAS two decades ago are too bright to be observed by IRS!  The
superior sensitivity of IRS will allow to detect dust in galaxies at
high redshift (\cite{brandl}) and cool faint nearby stars. The small
sizes of the slits, which are similar to the size of the point spread
function, and the availability of the spectral-mapping mode will be
powerful tools in tracing weak mid-infrared lines over a wide range of
ISM conditions. This improved spatial resolution should permit a
better comparison between the mid-infrared and the optical/radio
properties of the ISM and result to a better understanding the
underlying physical processes.

\section{Conclusion}

More information on the IRS and on upcoming deadlines for observing
opportunities with SIRTF can be found at the following web sites:\\
\\
http://www.astro.cornell.edu/SIRTF\\
http://sirtf.caltech.edu

\appendix


\begin{thebibliography}{}


\bibitem[\protect\astroncite{Fanson et al.}{1998}]{fanson}
Fanson J.L., Fazio G., Houck J.R., et al., 1998, Proc. SPIE Vol. {\bf
3356}, 478--491.

\bibitem[\protect\astroncite{Houck et al.}{1995}]{houck} 
Houck J.R., van Cleve J.E., 1995, Proc. SPIE Vol. {\bf 2475}, 456--463.

\bibitem[\protect\astroncite{Brandl et al.}{2000}]{brandl} 
Brandl, J.R. Charmandaris, V., Uchida, K.,  Houck, J.E., 2000,
in Proceedings of ``ISO Surveys of a Dusty Universe'', (astro-ph/9912216)

\bibitem[\protect\astroncite{Verstraete et al.}{2000}]{verstraete} 
L. Verstraete L., Pech C., Moutou C., Sellgren K., L\'eger A., 2000, (this
volume)

\end{thebibliography}
\end{document}